# THE IMPACT OF COVID-19 AND POLICY RESPONSES ON AUSTRALIAN INCOME DISTRIBUTION AND POVERTY

Jinjing Li[1*], Yogi Vidyattama[*], Hai Anh La[*], Riyana Miranti[*], Denisa M. Sologon[†]

[*]NATSEM, University of Canberra
[†]Luxembourg Institute of Socio-Economic Research

**ABSTRACT**

This paper undertakes a near real-time analysis of the income distribution effects of the COVID-19 crisis in Australia to understand the ongoing changes in the income distribution as well as the impact of policy responses. By semi-parametrically combining incomplete observed data from three different sources, namely, the Monthly Longitudinal Labour Force Survey, the Survey of Income and Housing and the administrative payroll data, we estimate the impact of COVID-19 and the associated policy responses on the Australian income distribution between February and June 2020, covering the immediate periods before and after the initial outbreak. Our results suggest that despite the growth in unemployment, the Gini of the equalised disposable income inequality dropped by nearly 0.03 point since February. The reduction is because of the additional wage subsidies and welfare supports offered as part of the policy response, offsetting a potential surge in income inequality. Additionally, the poverty rate, which could have been doubled in the absence of the government response, also reduced by 3 to 4 percentage points. The result shows the effectiveness of temporary policy measures in maintaining both the living standards and the level of income inequality. However, the heavy reliance on the support measures raises the possibility that the changes in the income distribution may be reversed and even substantially worsened off should the measures be withdrawn.

---

[1] Corresponding author. Email: jinjing.li@canberra.edu.au

# I. INTRODUCTION

The COVID-19 pandemic and the associated government response led to a significant shift in the norm of social and economic activities. Many governments worldwide have imposed various measures in the hope of containing the outbreak. Both the pandemic and the policy measures have wide-reaching and highly asymmetric implications on many aspects of life in most countries (Baker et al., 2020; Bonaccorsi et al., 2020; Mckibbin & Fernando, 2020). In Australia, all international travel has been banned since 20 March, and the government imposed extended lockdowns of its businesses in periods with rapid virus spread[2]. The border closure and lockdown, without a doubt, have had a significant impact on the economy, reflected in elevated unemployment levels. According to the National Skill Commission (2020), 17 of the 19 industries recorded reductions in employee jobs between mid-March and the end of May.

The government announced several policy initiatives to minimise job losses and support families, particularly the ones relying on welfare payments. The measures include two payments of $750 to certain welfare recipients, temporarily doubling the base payment rate for unemployment and other selected benefits, a wage subsidy programme which provides affected businesses with a flat rate of $1500 a fortnight per eligible employee, and a temporary increase in child care subsidy.

While the government has collected statistics to demonstrate the impact of the COVID-19 at the aggregate level such as industry, occupation, due to the absence of any compressive household survey, it is unclear how the pandemic has affected the living standard of the households and the level of income inequality. This information, however, is critical in understanding which sections of the population suffered the most from income fluctuations, which in turn supports formulating appropriate policy responses targeted to help those who were most affected.

This paper aim is to undertake a near real-time analysis of the income distribution effects of the COVID-19 crisis in Australia to understand the ongoing changes in the income distribution in order to support appropriate policy design. We propose a calibrated income distribution estimation based upon: (i) actual data on changes in employment (both at the extensive and intensive margin) drawn from the up-to-date Monthly Longitudinal Labour Force Survey; (ii) administrative data on the changes of the earnings at the industry and age group level drawn from the Weekly Payroll Jobs and Wages in Australia; and (iii) the Survey of Income and Housing which is a comprehensive household income survey collected in 2017-18. While most countries are in a situation where there is lack of timely data required to understand the distributional impact of the crisis, we adopt a novel approach that builds upon the latest developments in nowcasting the income distribution using non-parametric tools and this unique combination of datasets available in Australia. The approach enables us to plausibly simulate the counterfactual distributions of income under the impact of the crisis.

The paper contributes to the literature in three ways. First, we show that our proposed method can construct a near real-time household sample for income analyses from partially observed data, capturing the short-term economic changes and resembling the official statistics. Second, we provide the first estimates of the effects of the COVID-19 on the income distribution in Australia and show evidence on the heterogeneous effects based on individual-level data. Third, we estimate the contribution of the government response in stabilising the income distribution, showing the importance and the level of reliance on the temporary measures.

---

[2] For a complete timeline of Coronavirus outbreak in Australia and the government response, please refer to the News update archive from the Australian government website (Australia Government, 2020).

## II. STATE-OF-THE-ART IN NOWCASTING AND ITS NEED IN TIMES OF CRISIS

Modelling the socio-economic consequences of a shock is vital to inform policy decision-making (Bok et al., 2018). The COVID-19 pandemic and the subsequent restriction to control its spread have not only caused a downturn in the economic performance marked by negative GDP growth (Mckibbin & Fernando, 2020), but is likely to affect the levels of poverty and income inequality (Bonaccorsi et al., 2020). Given how widespread COVID-19 is globally and how restrictive social measures are, it is crucial to understand the impact of the crisis on inequality and poverty in a timely manner in order to formulate appropriate policy responses.

Historically, major economic shocks tend to induce changes in income inequality. In the 1997-1999 Asian Financial crisis, South Korea was one of the worst-hit economies among developed countries with contracted GDP, rising unemployment and sharply worsened income inequality (Cheong, 2001). The 2008 global financial crisis has severely affected all EU Member States with a plunge in real GDP ranging from 5 per cent in Cyprus to 40 per cent in Latvia (De Beer, 2012). The unemployment rose rapidly, and the governments in the EU introduced fiscal interventions with unemployment and welfare benefits. Nevertheless, income inequality increased in many countries. Although most EU countries started to recover in the new decade, many governments have accumulated substantial debt (Matsaganis & Leventi, 2014). This highlights the need for the government to come up with a well-targeted policy as well as a careful allocation of resources to avoid plunging into austerity after the initial stimulus (Matsaganis & Leventi, 2014).

While members of the society can feel the changes in the economic activities almost immediately after the spread of COVID-19, the statistics to understand the current state of the economy are often limited and will only be available with a significant time lag (Chetty et al., 2020). Detailed household information is rarely available in a timely manner given that most household surveys are collected annually (or less frequent) due to the substantial cost involved. Thus, there is always a non-negligible gap between data collection and its release for research, which restricts the analysis of a sudden event, such as the spread of COVID-19.

To overcome this gap, an increasing number of studies resorted to adapting the *Nowcasting* technique to represent statistics of the present state, the recent past and the near future by updating the latest collected data with up-to-date external statistics or controls (Bańbura et al., 2013). While the idea originated in meteorology, economists have adapted the technique to estimate timely economic indicators. Giannone et al. (2008) have used 200 macroeconomic indicators for the US economy such as industrial production, employment, financial variables, prices, wages, money and credit aggregates, surveys from other sources, and other conjunctural measures to nowcast GDP growth. They highlighted the importance of indicators that are updated to reflect the present time, such as consumer confidence index, unemployment insurance weekly claims report, interest rates, foreign exchange rates, stock exchange and S & P indices. More recently, nowcasting techniques have been applied to produce current estimates of poverty and income inequality. Some research use inflation indexation, adjusting proportionally the industry-specific employment rates combined with a tax-benefit simulator to evaluate the policy impact of various tax-benefit rules (Navicke et al., 2014).

A few European countries have analysed the impact of COVID-19 based on the Nowcast approach. Examples include Beirne et al. (2020) and O'Donoghue et al. (2020) for Ireland, Figari & Fiorio (2020) for Italy, Bronka et al. (2020) and Brewer & Gardiner (2020) for the UK. In the case of UK, earnings subsidies were found to cushion household income across the entire distribution, providing the primary insurance mechanism against the adverse income shock; existing tax-benefit rules were found to complement this scheme in providing social protection. In the case of Ireland, O'Donoghue et al (2020) find that the crisis-induced income-support policy responses (wage subsidies, pandemic-related unemployment and sickness benefits) combined with existing progressive elements of the tax-benefit system were effective in counteracting the increase in income inequality in the early months of the pandemic.

The existing nowcast approaches used, however, rely heavily on parametric assumptions regarding the changes in the labour market. The analyses tend to assume the changes both in the extensive and intensive margin are either stochastic or based on the parametric regressions derived from the pre-crisis household survey. The implicit assumption of the labour market transition is random or remains stable in the event of the crisis may not always hold, especially when shocks are expected to have highly heterogeneous impacts across the population. To address this limit, this paper proposes a semi-parametric nowcast approach combining multiple incomplete microdata as well as administrative statistics for estimations of the income distribution, exploiting the availabilities of both the comprehensive household survey as well as the shorter monthly labour force survey in Australia.

### III. DATA

As a comprehensive household survey is not available to capture the pre- and post- changes in the income distribution, we reconstruct the income distribution profiles by combining three different datasets that describe the evolution of the labour market in Australia. Specifically, this includes the monthly longitudinal labour force survey (LLFS), the biannual survey of income and housing (SIH), and the weekly payroll statistics. All datasets are collected by the Australian Bureau of Statistics.

The Longitudinal Labour Force Survey (LLFS) provides details on geography, demography (e.g., age, sex, educational attainment, family characteristics), the labour market participation (e.g., hours worked, industry, occupation, details of the last job) and the transitions into and out of the employment of Australians (Australian Bureau of Statistics, 2020b). The data however, does not contain any income information. The current LLFS consists of over 400 waves since 1982. Each monthly survey collects around 50,000 observations sampled from private dwellings and is representative of the population aged 15 or older. Given the rotation design of the sampling, 80-85% of the observations are retained in the successive wave.

The study uses the waves between February 2020 and June 2020 (5 waves) which covers the periods immediately before and after the COVID-19 starting to spread in Australia. The selection of the data period allows us to capture the effect of the COVID-19 progression in Australia.

The second data source we use is the Survey of Income and Housing (SIH) 2017-18, which collects data on demographics (e.g., sex, age, marital status), housing, labour force characteristics, education, income, assets and childcare (Australian Bureau of Statistics, 2019). At the individual level, the SIH provides detailed information of those living in private dwellings on their labour force status, the number of jobs currently held, weekly hours worked, duration of unemployment as well as detailed income information. Our study uses the 2017-18 version of the SIH, which includes the final sample of 14,060 households and provides detailed industry and occupation information that enables us to match with the changes in the monthly LLFS to examine the likely consequence of the income distribution due to COVID-19.

To refine the modelling of the heterogeneous shock which affects earnings across different industries, occupations, and demographic profiles, we further incorporate a longitudinal dataset of Weekly Payroll Jobs and Wages which is sourced from the administrative data, covering businesses with Single Touch Payroll (STP)-enabled payroll or accounting software and is provided to the Australian Taxation Office (ATO). STP generally covers all employers in Australia since July 2019 unless an exemption is granted. This data includes changes in payroll jobs, changes in total wages paid, and changes in average weekly wages per job (Australian Bureau of Statistics, 2020a). This data is also linked with other administrative data from the Australian taxation system to determine additional classification attributes (e.g. age and sex of employees).

The three datasets can be understood as the different projections of the socio-economic profiles of Australia, although each alone does not provide sufficient information for our analyses. The combination of the three sources of data can offer a more complete picture of the socio-economic changes. Additionally, as the information related to the labour market change, the primary venue of the shocks of COVID-19, is already captured in the data, there is less of a need to specific parametric equations to model to the evolution of the labour market, avoiding many restrictive assumptions. The use of multiple unit-record datasets enables us to reconstruct the socio-economic profile semi-parametrically to reflect the observed changes due to COVID-19, unlike most of the existing nowcast approach in the literature.

## IV. METHODOLOGY

*General Estimation Strategy*

Given the time since the COVID-19 outbreak in Australia, we focus on the short-term change in income distribution. We also assume factors other than COVID-19 such as demographic change and other unpredicted events are negligible in reshaping the income distribution in the immediate periods after the lockdown. The changes of the income distribution can therefore be decomposed into two main parts, with one being the change induced by the income shock (part A) and the other due to the government policy response (part B):

$$\Delta I = I(p_1, y_1) - I(p_0, y_0)$$
$$= \underbrace{I(p_0, y_1^*) - I(p_0, y_0)}_{\text{Income Shock Effect (A)}} + \underbrace{I(p_1, y_1) - I(p_0, y_1^*)}_{\text{Policy Response Effect (B)}} \quad (1)$$

where $I$ is an outcome measure calculated from the entire income distribution (e.g., average disposable income, Gini index, poverty rate etc.), $p$ is the government COVID-19 response policies and $y$ is the distribution of the gross market income. The subscript 0 refers to the pre-COVID-19 baseline period, which is February 2020, and the subscript 1 refers to the periods after the initial spread of COVID-19, ranging between April and June 2020[3]. We use $y_1^*$ to denote the unobserved gross market income distribution sans the policy intervention after the initial spread of COVID-19.

Equation (1) resembles a similar decomposition framework which quantifies the contribution of the specific component to inequality changes between two points in time using a counterfactual simulation approach (Bargain et al., 2017; Bargain & Callan, 2010). Part A in Equation (1) captures the changes in income distribution due to the change in the market income. As COVID-19 dominates the income shock between Feb 2020 and June 2020, we assume that most (if not all) changes in Part A can be attributed to COVID-19. Part B captures the policy changes and their effects on income distribution. As all changes in the policies are part of the government response to COVID-19, the effect of B is also a measure of the effectiveness of the policy response in stabilising the market income shock via welfare policies. It should be noted that Equation (1) does not incorporate the interaction term of the two major components as the policy response can only be meaningfully interpreted conditional on the existence of COVID-19. Therefore, a sequential decomposition framework becomes more suitable in this particular analysis. The outcome measures in this paper include the mean and the Gini of the gross market income as well as the equivalised disposable income. Additionally, we decompose the changes in the poverty rate by holding the poverty line constant at the pre-COVID-19 level.

---

[3] As March can be considered as a transitory month in terms of both policy measures and the propagation of the economic shocks, we generally do not interpret the results from March.

The estimation of the Equation (1) replies on the estimation of three income distributions[4], namely the pre-COVID-19 income distribution $I(p_0, y_0)$, the COVID-19 income distribution $I(p_1, y_1)$, and the counterfactual COVID-19 income distribution without any policy response $I(p_0, y_1^*)$. We use the samples from February to estimate $I(p_0, y_0)$ and the samples from a month later than March to estimate $I(p_1, y_1)$. The more challenging part of the decomposition is the estimation of $I(p_0, y_1^*)$, where we need to simulate the likely outcome should there be no welfare policy intervention ($p_0$) amid the COVID-19 market income shock ($y_1$). The estimation can be difficult due to the nature of the shock as well as the complex behaviour response associated with it. More specifically, we do not know the demand of the labour market if the government does not provide the subsidy that affects approximately 3.5 million jobs (Australian Treasury, 2020). To accommodate the possible behavioural and market responses, instead of providing a point estimate with a strong assumption, we estimate the upper and the lower bound of $I(p_0, y_1^*)$ using two extreme estimates:

- Lower bound estimate: we assume no job will be lost should the government withdraw the additional subsidies to the employers;
- Upper bound estimate: we assume that all subsidised jobs would be lost should there be no subsidy.

Although neither estimate above is realistic, the range is sufficiently large to provide a robust estimate of the term A and B in Equation (1). The difference between the two estimates reflects the uncertainty in the behavioural response. It is most likely that the most realistic counterfactual resides somewhere in between these two extreme estimates.

*Constructing Demographics, Labour Market and Income Distributions*

To capture the changes in the demographic and employment while retaining the income information, we estimate a set of the new weights of the latest Survey of Income and Housing (SIH) to mimic the demographic and employment patterns observed in the monthly Longitudinal Labour Force Surveys (LLFS), both before and after the initial spread of COVID-19. We follow the semi-parametric method by DiNardo, Fortin and Lemieux (1996) to adjust the weight of each observation in the 2017-18 SIH with an estimated ratio. Specifically, the weight of an individual $i$ from SIH ($w_{i,\text{SIH}}$) is updated using the Bayesian rule:

$$w_{i,\text{SIH}\to\text{LLFS}} = w_{i,\text{SIH}} \frac{\Pr(X_i|\text{LLFS})}{\Pr(X_i|\text{SIH})} = w_{i,\text{SIH}} \frac{\Pr(\text{LLFS}|X_i)}{\Pr(\text{SIH}|X_i)} \gamma \quad (2)$$

where $\Pr(X_i|\text{LLFS})$ and $\Pr(X_i|\text{SIH})$ are the conditional probabilities of observing the characteristics $X_i$ in the LLFS and SIH dataset respectively; $\Pr(\text{LLFS}|X_i)$ and $\Pr(\text{SIH}|X_i)$ are the probabilities of observation with characteristics $X_i$ drawn from LLFS and SIH, respectively. The characteristics vector $X$ contains:

- demographic characteristics (age, age squared, marital status, gender, oversea born);
- employment characteristics (industry, occupation, usual hours of working, number of jobs, duration of unemployment, and their interactions with gender and age);
- household characteristics (number of young kids, size of the household and state in which the household resides).

---

[4] The estimation of (1) can also be done using a different path of the decomposition, where $I(p_1, y_0)$ can be used as a counterfactual instead of $I(p_0, y_1^*)$. This is sometimes referred as the path dependency issue. However, the measure $I(p_1, y_0)$ does not have a meaningful interpretation as the COVID-19 policy response will exist only when COVID-19 takes place. We therefore adopt the sequential decomposition approach as in the main text.

The ratio $\gamma$ can be interpreted as the prior, which is the relative ratio of the dataset size. To ensure the weighted total population in our model matches the expected national population, we adjust the ratio slightly to reflect the population growth between the time of the SIH collection and 2020. The implicit assumption behind our approach is that observed changes in the intensive and extensive margins of the labour market can be effectively described through the semi-parametric process, and the income distribution, conditional on the observed characteristics including the working hours and the employment status remains largely stable other than the wage level adjustment reflected by the payroll information. This also mimics the short-term inelasticity of the conditional wage rate due to contracts and the labour law.

Besides the distributional adjustment, all income items including wage, investment income, and certain expenses are also indexed to account for the income growth between the time of SIH collection and the month of the analyses. Specifically:

- The wage income is indexed using ABS-published Average Weekly Earnings (AWE) growth rate between 2018 and February 2020. Since March 2020, we use the ABS weekly payroll information to reflect the heterogeneous wage shock across different industries and age groups. The monthly wage growth (or decline) is estimated from the changes in the administrative payroll statistics by industry and age group.
- The investment income is assumed to grow at a real return of 2.5% between 2018 and early 2020. To reflect the volatility of COVID-19 on the financial market, we use the default portfolio performance of the largest pension fund in Australia (Australian Super, 2020) to index the investment income between February and June 2020. Figure A in the Appendix plots the portfolio performance since January 2020.
- The rest of the income and the childcare expenses are indexed using the Consumer Price Index (CPI). The childcare expenditure is used for the post-childcare disposable income estimation.

*Modelling Policy Responses to the COVID-19 Crisis*

In response to the COVID-induced economic shocks, the Australian government announced several temporary taxation and welfare benefit changes that directly affect the employment and the income level of the households, including a one-off payment for specific welfare recipients, increased unemployment and other benefits, the temporary free childcare support, and wage subsidies to employees via eligible employers.

In March 2020, the government announced a one-off economic support payment of $1500 to nearly existing recipients of specific welfare payment, including pensions, unemployment benefits (Jobseekers) and Family Tax Benefit (FTB). The majority of recipients received the first payments ($750) by mid-April, and the second payment ($750) was made starting from mid-July. Besides one-off payments, the government also introduced a temporary payment of $550 per fortnight, known as the Coronavirus Supplement, to eligible welfare payment recipients who receive unemployment benefit (Jobseeker), parenting payment, youth allowances or other payment. The unemployment benefit has also been doubled to $1,115.70 per fortnight while its eligibility criteria were relaxed as recipients do not need to fulfill their usual job searching obligations (as known as "mutual obligation"). Additionally, the taper rate, which determines the reduction of the payment when the partner's income exceeds the threshold, has also been temporarily changed (Centrelink, 2020). In the meantime, asset testing has also been suspended for selected benefits (Centrelink, 2020).

Along with the changes in direct welfare payments, the government also provides full subsidies to childcare services for approximately three months from the start of April to the end of June. The measure intends to help childcare providers stay open and keep employees in their jobs. The usual income and activity test for childcare subsidy were suspended during the period. The spending is projected at $1.6 billion in costs of the relief package over three months.

To estimate the impact of the policy changes, we use the latest version of an Australian tax-transfer model STINMOD+ to calculate household disposable income based on the corresponding tax and social transfer rules, including the aforementioned COVID-19 stimulus and the policy changes in direct welfare payment and childcare subsidy. The model covers all personal taxation and federally administered welfare payments and replicates the implementation of the social security system, incorporating elements such as income and asset testing. As the Australian welfare system is highly means-tested, and the vast majority of the eligible conditions do not depend on previous contributions or complex employment history, we can estimate the welfare payment with relatively high accuracy. As the take-up rate of means-tested benefits in Australia is generally high with a relatively low stigma (Mood, 2006), we assume full take-up in our simulation.

To reflect the relaxed job search requirement in the unemployment benefit, we assume all individuals losing their jobs since February are eligible for the unemployment benefit even if they are not actively seeking jobs. We use LLFS to estimate the propensity of that one was employed in any wave since February prior to being out of labour force with a standard probit model (see Table B in the Appendix for the model specification) and applied the coefficients to the SIH data to simulate the additional eligibility of unemployment benefit for those who are currently out of the labour force.

Additional to the increased welfare payment, the government also announced a wage subsidy package called 'JobKeeper', providing eligible employers with a flat rate of $1,500 AUD per fortnight per employee, irrespective of prior or current hours and earnings. To be eligible, employers have to show that their turnover has reduced by at least 50 percent for large firms and 30 percent for smaller firms during the pandemic (Australian Treasury, 2020). The flat payment rate means that many eligible part-time or long-term casual employees can earn more than their regular pay. A recent review from the Australian Treasury suggests that the JobKeeper program has covered over 920,000 organisations and around 3.5 million individuals during April and May 2020, with a total payment of exceeding $20 billion (Australian Treasury, 2020).

As we do not have the employer-employee data that contains the eligibility status of the Job Keeper program, we need to simulate who is among the 3.5 million recipients as reported by the administrative data. Because the wage subsidy is only given to employers with severe financial difficulties and the intention is to retain jobs that would otherwise be at risk of disappearance, we assume the likelihood of receiving the benefit can be approximated by the immediate risk of losing the job. We use a standard probit model (see Table C in the Appendix for the model specification) to estimate the one's propensity of losing the job in the next wave (month), conditional on being employed in LLFS and apply the conditional probability to the SIH data. A common alignment algorithm is used to ensure the simulation will have the exact number of recipients matching the administrative data (Li & O'Donoghue., 2014).

## V. Results

*Validation - Comparisons between the observed and modelled distribution*

Before discussing the results from the estimated income distribution, it is crucial to assess whether the proposed method can construct a satisfactory distribution. For an evolving income distribution affected by a sudden shock on the labour market, one would expect that the estimated distribution to be sensitive to changes in the labour market conditions, whereas the general demographic distribution should remain mostly constant. An excessive deviation from the baseline demographic profile means that the reweighting process struggles to adjust the weight using observations available in the original survey data.

Table 1 describes the main demographic characteristics of the original SIH dataset, the up to date LLFS dataset and the constructed dataset based on the semi-parametric reweighting. As shown, the result from the nowcast model generally falls between the two datasets, suggesting a stable demographic profile.

Table 1 Comparison between modelled and observed demographic profile (Feb 2020)

| Variable | SIH | LLFS | Modelled |
|---|---|---|---|
| Proportion of population between 15-24 | 15.7% | 16.0% | 15.7% |
| Proportion of population between 25-64 | 65.7% | 66.6% | 65.3% |
| Proportion of population 65+ | 18.6% | 17.4% | 19.0% |
| Proportion of Male | 49.0% | 49.4% | 49.2% |
| Proportion of socially married | 61.0% | 59.8% | 59.5% |
| Australian born | 65.9% | 67.1% | 67.0% |
| Education bachelor or higher | 27.8% | 29.0% | 28.6% |
| Number of children under 15 in the household | 0.52 | 0.55 | 0.54 |

Table 2 compares the changes in the simulated unemployment rate versus the official unemployment rate published by the Australian Statistical Bureau between February 2020 and June 2020. The model result, especially, the longitudinal change between March and April indicates that the method can capture the changes in the employment distribution. In the baseline month (February), the difference between the simulated and the official unemployment rate is less than 0.1 percentage point. In other months, the differences are generally small with the maximum discrepancy observed in June. Results from a sub-population, e.g. youth unemployment, is less accurate due to the reduced number of observations although the results are generally comparable. As each month's distribution is estimated separately, variations also come from statistical uncertainties such as sampling errors and the use of the random numbers in the simulation. Throughout the paper, we report all estimates between February and June (with April to June as the COVID-19 affected period) to more accurately assess the robustness of the results.

Table 2 Unemployment Rate compared with official figures

|  | **Unemployment Rate** | | **Youth Unemployment** | |
|---|---|---|---|---|
|  | Modelled | Official | Modelled | Official |
| **Feb** | 5.45% | 5.52% | 13.14% | 13.18% |
| **Mar** | 5.69% | 5.57% | 13.74% | 12.64% |
| **Apr** | 6.66% | 6.43% | 15.21% | 14.21% |
| **May** | 6.90% | 6.92% | 15.20% | 15.19% |
| **Jun** | 8.66% | 7.25% | 18.88% | 15.64% |

*Changes in Labour Market and the Income Distribution*

Table 3 reports the estimated propensity of dropping out of employment conditional on being employed in the previous period across income quintiles. Two patterns become apparent. First, the likelihood of dropping out generally falls as income increases, even at the height of the COVID-19 lockdown. This pattern echoes what has been observed in other crises caused by external shocks, namely that job security is affected differently across the income distribution (Baldacci et al., 2002). Second, the propensity of dropping out of employment exhibits an increasing trend between February and April 2020, followed by a decline in May and June 2020. The change in the probability also mirrors the intensity of the COVID-19 shocks over time: the main effects are observed in March-May, followed by a stabilising labour market trend starting in May 2020. This bounce-back pattern may also reflect that the government interventions that have commenced in March 2020 started to have an impact. Similar to previous crises, the COVID-19 shock also displays a differential impact across industries and income levels. Those employed but in the lowest income quintile were the hardest hit by losing their jobs in April 2020. These include those who work in the Arts and Recreation Services and Accommodation and Food Services sectors, the two most contracted sectors between February and May 2020 (See Table D in Appendix). Both industries suffered due to the impact of border closure and travel or mobility restrictions.

Table 3 Estimated propensity of dropping out of employment, conditional on being employed in the previous month, by market income quintile (wage and business income).

|  | Q1 | Q2 | Q3 | Q4 | Q5 |
|---|---|---|---|---|---|
| **February** | 2.66% | 1.47% | 1.08% | 0.85% | 0.71% |
| **March** | 6.87% | 3.00% | 2.46% | 2.13% | 2.12% |
| **April** | 14.17% | 6.13% | 4.18% | 3.12% | 2.58% |
| **May** | 10.31% | 4.58% | 3.07% | 2.27% | 1.78% |
| **June** | 6.97% | 3.19% | 2.45% | 2.03% | 1.84% |

Table 4 reports the estimated average monthly household gross market income by pre-COVID quintiles of equivalised household disposable income in dollar amount and the percentage change compared with the pre-COVID period, taking into account the observed changes in the labour force participation as well as working hours. The market income change shows the extent of the COVID-19 shock as well as the impact of the government wage subsidy program (JobKeeper) in pre-tax earnings. In the pre-COVID period, the estimates show the gross household incomes are mostly stable. In May 2020, gross market income increased for most quintiles, with over 20% for Q1. Q5 is the only exception, where we observe an initial decline of the gross income in April but a gradual recovery since then.

The increase in gross market incomes between April and June in the lower quintiles can largely be attributed to the wage subsidy scheme and the increased welfare benefit as shown by the estimates without policy responses. In the lower bound estimates, where we assume that all jobs which are currently supported by the JobKeeper and the welfare support schemes are gone, a significant decline in gross earning is expected throughout the distribution. In the upper bound estimates, where no JobKeeper supported jobs are lost, the changes in gross incomes are generally less than a few percentage points. The actual impact of the wage subsidy on labour demand is likely in between those two extreme estimates. The difference between the nowcast estimates and the estimates without the policy change shows the substantial impact of the policy change, especially for households in Q1 and Q2.

Table 4 Estimated household monthly market income (labour, business, investment) by pre-COVID equivalised household disposable income quintiles

|  | Monthly Gross Income (A$) | | | | | % Change Compared to February | | | | |
| --- | --- | --- | --- | --- | --- | --- | --- | --- | --- | --- |
|  | Q1 | Q2 | Q3 | Q4 | Q5 | Q1 | Q2 | Q3 | Q4 | Q5 |
| **Before COVID-19** | | | | | | | | | | |
| February | 897 | 3508 | 7563 | 11749 | 22635 | | | | | |
| March | 889 | 3474 | 7548 | 11738 | 22660 | -0.9% | -1.0% | -0.2% | -0.1% | 0.1% |
| **Nowcast Model** | | | | | | | | | | |
| April | 1098 | 3764 | 7895 | 12070 | 22445 | 22.4% | 7.3% | 4.4% | 2.7% | -0.8% |
| May | 1110 | 3801 | 7941 | 12112 | 22622 | 23.7% | 8.4% | 5.0% | 3.1% | -0.1% |
| June | 1069 | 3641 | 7879 | 12092 | 22757 | 19.2% | 3.8% | 4.2% | 2.9% | 0.5% |
| **Without policy response (lower estimates)** | | | | | | | | | | |
| April | 730 | 2743 | 6042 | 9326 | 19127 | -18.6% | -21.8% | -20.1% | -20.6% | -15.5% |
| May | 723 | 2685 | 6001 | 9267 | 19224 | -19.4% | -23.5% | -20.7% | -21.1% | -15.1% |
| June | 708 | 2608 | 5916 | 9129 | 18818 | -21.1% | -25.7% | -21.8% | -22.3% | -16.9% |
| **Without policy response (upper estimates)** | | | | | | | | | | |
| April | 888 | 3457 | 7561 | 11754 | 22277 | -0.9% | -1.5% | 0.0% | 0.0% | -1.6% |
| May | 903 | 3479 | 7609 | 11804 | 22455 | 0.7% | -0.8% | 0.6% | 0.5% | -0.8% |
| June | 869 | 3379 | 7570 | 11806 | 22591 | -3.2% | -3.7% | 0.1% | 0.5% | -0.2% |

Following the decomposition framework as set out in the methodology section, we estimate the contributions from the income shock effect and the government policy response to the overall changes in the income. Table 5 reports the estimates and the range derived from the counterfactuals. The term "Low" is used in the table to indicate the lower bound estimates and the term "High" for the upper bound estimates. As shown, the income shock effect, which corresponds to term A in Equation (1), contributes to a maximum of a hundred dollars decline to the gross income for the lower-income household to up to a few thousand dollars for the high-income households. The policy response effect, which corresponds to the term B in the Equation (1), generally negates the decline induced by the income shock effect and contributes to an increase of a few hundred to a few thousand dollars a month to household market income. Overall, the income shock due to COVID-19 contributes a maximum of 15-25.7% to the decline in the gross income depending on the quintile, and the policy response contributes up to 43.1% of the pre-COVID gross income for households in Q1 and up to 30% for households in other quintiles. The net effect is what we observed in Table 4, with the impact being highly heterogeneous across quintiles.

Table 5 Estimated contributions of COVID-19 income shock effect and policy response effect to changes in monthly household market income by pre-COVID equivalised disposable income quintiles

|  | Q1 | | Q2 | | Q3 | | Q4 | | Q5 | |
| --- | --- | --- | --- | --- | --- | --- | --- | --- | --- | --- |
|  | Low | High | Low | High | Low | High | Low | High | Low | High |
| **Income Shock Effect (A$)** | | | | | | | | | | |
| April | -167 | -8 | -764 | -51 | -1522 | -2 | -2423 | 5 | -3508 | -358 |
| May | -174 | 7 | -823 | -29 | -1562 | 45 | -2483 | 55 | -3411 | -180 |
| June | -189 | -28 | -900 | -129 | -1648 | 7 | -2620 | 57 | -3817 | -43 |
| **Policy Response Effect (A$)** | | | | | | | | | | |
| April | 210 | 368 | 308 | 1021 | 333 | 1853 | 315 | 2743 | 168 | 3318 |
| May | 206 | 387 | 322 | 1116 | 332 | 1940 | 309 | 2846 | 167 | 3398 |
| June | 201 | 361 | 262 | 1033 | 309 | 1963 | 287 | 2963 | 166 | 3939 |
| **Income Shock Effect (Percentage of pre-COVID income)** | | | | | | | | | | |
| April | -18.6% | -0.9% | -21.8% | -1.5% | -20.1% | 0.0% | -20.6% | 0.0% | -15.5% | -1.6% |
| May | -19.4% | 0.7% | -23.5% | -0.8% | -20.7% | 0.6% | -21.1% | 0.5% | -15.1% | -0.8% |
| June | -21.1% | -3.2% | -25.7% | -3.7% | -21.8% | 0.1% | -22.3% | 0.5% | -16.9% | -0.2% |
| **Policy Response Effect (Percentage of pre-COVID income)** | | | | | | | | | | |
| April | 23.4% | 41.0% | 8.8% | 29.1% | 4.4% | 24.5% | 2.7% | 23.4% | 0.7% | 14.7% |
| May | 23.0% | 43.1% | 9.2% | 31.8% | 4.4% | 25.6% | 2.6% | 24.2% | 0.7% | 15.0% |
| June | 22.4% | 40.3% | 7.5% | 29.5% | 4.1% | 26.0% | 2.4% | 25.2% | 0.7% | 17.4% |

Table 6 reports the changes in equivalised household disposable income after childcare cost by pre-COVID quintiles, in absolute amounts and relative changes. We deduct the childcare cost from the disposable income as consumption of childcare services changed significantly due to COVID-19 and the government policy-response. In general, we observe a similar pattern as in Tables 4. The largest changes in equivalised disposable income levels are found in Q1 and Q2, with an increase of around 24% and 9% respectively. It should be noted that while the analysis shows an average improvement in the disposable income in the lower quintiles, it does not suggest all families are better off. The heterogeneity of shocks means that some individuals may be better off than the group average, while others could be significantly worse off, especially those who lost their jobs.

The impact of COVID-19 outbreak would have been greater if there had been no policy response. Disposable income could have contracted by 3-4% for Q1 and by around 10% for Q2 in the worst-case scenario. Due to the inability to access means-tested welfare payment for the higher income groups, Q3-Q5 would have suffered more percentage-wise due to job losses, although they are still better off in absolute term. Across all results, the variations over time between April and June tend to be small. As each month is estimated independently, the stability of the numeric results is a welcoming sign, suggesting the robustness of the results derived from the semi-parametric approach.

Table 6 Estimated monthly equivalised household disposable income by pre-COVID equivalised disposable income quintiles

| | Monthly Disposable Income (A$) | | | | | % Change Compared to February | | | | |
|---|---|---|---|---|---|---|---|---|---|---|
| | Q1 | Q2 | Q3 | Q4 | Q5 | Q1 | Q2 | Q3 | Q4 | Q5 |
| **Before COVID-19** | | | | | | | | | | |
| February | 1872 | 2826 | 4006 | 5541 | 9997 | | | | | |
| March | 1864 | 2812 | 4006 | 5557 | 9922 | -0.5% | -0.5% | 0.0% | 0.3% | -0.7% |
| **Nowcast Model** | | | | | | | | | | |
| April | 2311 | 3087 | 4191 | 5670 | 9940 | 23.5% | 9.2% | 4.6% | 2.3% | -0.6% |
| May | 2322 | 3083 | 4162 | 5621 | 9866 | 24.0% | 9.1% | 3.9% | 1.4% | -1.3% |
| June | 2322 | 3078 | 4206 | 5694 | 10041 | 24.0% | 8.9% | 5.0% | 2.7% | 0.4% |
| **Without policy response (lower estimates)** | | | | | | | | | | |
| April | 1808 | 2542 | 3380 | 4557 | 8595 | -3.4% | -10.1% | -15.6% | -17.8% | -14.0% |
| May | 1800 | 2514 | 3334 | 4475 | 8490 | -3.8% | -11.1% | -16.8% | -19.2% | -15.1% |
| June | 1816 | 2539 | 3379 | 4485 | 8440 | -3.0% | -10.2% | -15.7% | -19.1% | -15.6% |
| **Without policy response (upper estimates)** | | | | | | | | | | |
| April | 1881 | 2846 | 4049 | 5606 | 9934 | 0.5% | 0.7% | 1.1% | 1.2% | -0.6% |
| May | 1880 | 2838 | 4021 | 5562 | 9857 | 0.4% | 0.4% | 0.4% | 0.4% | -1.4% |
| June | 1888 | 2840 | 4060 | 5627 | 10020 | 0.8% | 0.5% | 1.4% | 1.6% | 0.2% |

Table 7 Estimated contributions of COVID-19 income shock effect and policy response effect to changes in equivalised disposable income by pre-COVID equivalised disposable income quintiles

| | Q1 | | Q2 | | Q3 | | Q4 | | Q5 | |
|---|---|---|---|---|---|---|---|---|---|---|
| | Low | High | Low | High | Low | High | Low | High | Low | High |
| **Income Shock Effect (A$)** | | | | | | | | | | |
| April | -64 | 9 | -284 | 20 | -626 | 43 | -984 | 65 | -1401 | -62 |
| May | -72 | 7 | -312 | 12 | -672 | 15 | -1066 | 21 | -1507 | -139 |
| June | -56 | 16 | -287 | 14 | -627 | 54 | -1056 | 86 | -1556 | 23 |
| **Policy Response Effect (A$)** | | | | | | | | | | |
| April | 430 | 503 | 241 | 545 | 142 | 811 | 64 | 1113 | 5 | 1344 |
| May | 442 | 521 | 245 | 569 | 141 | 828 | 59 | 1146 | 8 | 1376 |
| June | 434 | 506 | 238 | 539 | 146 | 827 | 66 | 1208 | 21 | 1600 |
| **Income Shock Effect (Percentage of pre-COVID income)** | | | | | | | | | | |
| April | -3.4% | 0.5% | -10.1% | 0.7% | -15.6% | 1.1% | -17.8% | 1.2% | -14.0% | -0.6% |
| May | -3.8% | 0.4% | -11.1% | 0.4% | -16.8% | 0.4% | -19.2% | 0.4% | -15.1% | -1.4% |
| June | -3.0% | 0.8% | -10.2% | 0.5% | -15.7% | 1.4% | -19.1% | 1.6% | -15.6% | 0.2% |
| **Policy Response Effect (Percentage of pre-COVID income)** | | | | | | | | | | |
| April | 23.0% | 26.9% | 8.5% | 19.3% | 3.5% | 20.2% | 1.2% | 20.1% | 0.1% | 13.4% |
| May | 23.6% | 27.9% | 8.7% | 20.2% | 3.5% | 20.7% | 1.1% | 20.7% | 0.1% | 13.8% |
| June | 23.2% | 27.0% | 8.4% | 19.1% | 3.6% | 20.6% | 1.2% | 21.8% | 0.2% | 16.0% |

Table 7 provides the estimates of the contributions from the COVID-19 and policy response to changes in disposable income. Compared with the estimates of the gross income as in Table 6, the absolute dollar amount change is generally lower, suggesting the automatic stabilising effect of the existing tax and benefit system. The equivalised disposable income level for Q1 households would only have minor changes with or without any direct policy response as existing welfare payment is an important source of income which is unaffected by the COVID-19 outbreak. Households in Q2 to Q5 would have experienced a larger reduction under a no policy response scenario. The income shock effect would account for up to 20% of the reduction of disposable income in the worst-case scenario. The policy response effect is a net positive for all quintiles, responsible for the increase in disposable incomes, as we observed in Table 6, with Q1 households benefiting the most.

Table 8 further breaks down the changes reported in Table 7 into the effect of the temporary free childcare, the effect of the change in gross incomes, which captures the employment stock as well as the wage subsidy, and the rest (welfare policy changes, as well as the interactions between the existing welfare system and the income shock). The change in disposable income is dominated by the increased level of welfare for Q1. Gross income change becomes a more important factor relative to others from Q2 onwards, becoming the primary driver for changes in disposable income for the wealthiest households. Free childcare has only a limited impact on the final disposable income, contributing around half a percentage point to the total change of the equivalised disposable income as it only affects the out-of-pocket portion of a service that is already heavily subsidised.

Table 8 Contributions of free childcare and gross income shock (including JobKeeper subsidy) on the estimated changes in equivalised household disposable income by equivalised household disposable income quintile (%)

|  |  | Q1 | Q2 | Q3 | Q4 | Q5 |
|---|---|---|---|---|---|---|
| **April** | Total Change | 23.5% | 9.2% | 4.6% | 2.3% | -0.6% |
|  | Free Childcare | 0.5% | 0.7% | 0.9% | 0.7% | 0.6% |
|  | Gross Income | 5.6% | 3.8% | 3.1% | 1.9% | -1.9% |
|  | Rest | 17.4% | 4.7% | 0.7% | -0.3% | 0.7% |
| **May** | Total Change | 24.0% | 9.1% | 3.9% | 1.4% | -1.3% |
|  | Free Childcare | 0.4% | 0.7% | 0.9% | 0.7% | 0.6% |
|  | Gross Income | 5.8% | 4.3% | 3.3% | 2.1% | -1.3% |
|  | Rest | 17.8% | 4.1% | -0.3% | -1.3% | -0.6% |
| **June** | Total Change | 24.0% | 8.9% | 5.0% | 2.7% | 0.4% |
|  | Free Childcare | 0.5% | 0.7% | 0.9% | 0.7% | 0.6% |
|  | Gross Income | 4.6% | 1.4% | 2.4% | 1.7% | -0.7% |
|  | Rest | 18.9% | 6.8% | 1.7% | 0.4% | 0.6% |

*Changes in Inequality*

This section examines how the changes observed across the distribution are reflected in the changes in the market income inequality (for everyone in age between 15 and 64) and disposable income inequality. The nowcast estimates in Table 9 suggest that gross income inequality has been relatively stable over time except for June, where there is almost a 0.02 point increase compared with February. However, when decomposing the change into the income shock effect and the policy response effect, it becomes clear that the COVID-19 income shock would cause an increase in gross income inequality ranging between 0.016 to 0.14 Gini points in April-June relative to February. The impact, however, is almost entirely negated by the policy effect due to the introduction of the wage subsidy. This finding is robust across both two extreme estimates. As the estimates include both positive and negative numbers, we do not use the term "lower" or "upper" estimates in this table to avoid confusions. Instead, the more pessimistic estimates where the absence of wage subsidy would lead to job losses of all supported jobs are labelled as "high impact", and the more optimistic estimates are labelled as "low impact". The two extreme estimates help to provide the bounds of the estimates for the policy and the income effects.

Table 9 Market Income (Wage, Business and Investment) Gini (Age 15-64)

|  | Feb | Mar | Apr | May | Jun |
| --- | --- | --- | --- | --- | --- |
| **Gini of Gross income** | 0.539 | 0.543 | 0.534 | 0.539 | 0.557 |
| **Changes relative to Feb** | - | 0.005 | -0.005 | 0.000 | 0.018 |
| **Income Shock Effect (Low impact)** | - | 0.005 | 0.016 | 0.020 | 0.036 |
| **Income Shock (High impact)** | - | 0.005 | 0.115 | 0.121 | 0.139 |
| **Policy Effect (Low impact)** | - | 0.000 | -0.020 | -0.020 | -0.018 |
| **Policy Effect (High impact)** | - | 0.000 | -0.120 | -0.121 | -0.121 |

Compared with the relatively stable gross market income inequality due to wage subsidies, disposable income inequality shows an apparent reduction as indicated in Table 10. Compared with February 2020, when the Gini was around 0.33, the Gini has been hovering between 0.302 to 0.305 since April. A drop of 0.03 Gini points in a few months is substantial given the annual change in disposable income inequality measures by the Gini index in Australia is usually smaller than 0.01 (Li et al., 2020). By comparison, the change in Gini between 2007 and 2009 was less than 0.012 over the two years. As expected, the policy contributed to a substantial reduction in income inequality, which dominates the rise induced by the income shock as evaluated by our low and high impact estimates. This downward change in income inequality is not unique to Australia. A similar policy impact during the COVID-19 crisis was found by O'Donoghue et al. (2020) and Beirne et al. (2020) in Ireland and Brewer and Tasseva (2020) in the UK. They found that the wage subsidy scheme cushioned family income across the distribution. Coupled with the rules embedded in the tax-benefits systems, it provided much-needed income protection.

Table 10 Equivalised Disposable Income Gini

|  | Feb | Mar | Apr | May | Jun |
| --- | --- | --- | --- | --- | --- |
| **Gini of Disposable Income** | 0.330 | 0.329 | 0.303 | 0.302 | 0.305 |
| **Changes vs Feb** | - | -0.001 | -0.027 | -0.027 | -0.024 |
| **Income Shock Effect (Low impact)** | - | -0.001 | -0.001 | -0.001 | 0.003 |
| **Income Shock (High impact)** | - | -0.001 | 0.042 | 0.046 | 0.048 |
| **Policy Effect (Low impact)** | - | 0.000 | -0.026 | -0.026 | -0.028 |
| **Policy Effect (High impact)** | - | 0.000 | -0.069 | -0.073 | -0.073 |

*Changes in Poverty*

As the poverty rate in Australia is often calculated after deducing housing costs, we defined a comparable poverty line which is 50% of the median equivalent income (OCED-modified equivalence scale) after the deduction of the housing cost and childcare cost. The poverty line is kept constant after February 2020 to enable the direct comparison before and after the initial COVID-19 shock. Our estimates in Table 11 suggest that the poverty rate has declined since the policy response to COVID-19 was implemented in April 2020. The poverty rate in February was around 13.5% which is consistent with the known poverty estimates in recent years (ACOSS, 2020). It has subsequently declined by 5 to 6 percentage points compared with February and March. This outcome is in contrast to the scenario of an absent policy response, where the poverty rate would jump by 12 percentage point compared with the pre-COVID period (in the upper estimates) or slightly increase during the same period (in the lower estimates case).

Table 11 After-housing poverty rate

|  | Feb | Mar | Apr | May | Jun |
|---|---|---|---|---|---|
| **Nowcast Model** | 13.5% | 13.8% | 8.0% | 8.1% | 8.2% |
| **w/o policy response (lower estimates)** | 13.5% | 13.8% | 13.8% | 14.1% | 14.6% |
| **w/o policy response (upper estimates)** | 13.5% | 13.8% | 25.8% | 27.4% | 28.2% |

## VI. CONCLUSIONS

Similar to other countries, COVID-19 has had a significant impact on social and economic activities in Australia. Using the combined data from the monthly longitudinal labour force survey, payroll information and the survey of income and housing, the paper propose a method to reconstruct the income distribution semi-parametrically from incomplete data and estimates the impact of the COVID-19 outbreak and the policy response on the income distribution.

While COVID-19 has increased unemployment, the wage subsidy initiative from the government has been effective in increasing market incomes for households in the lowest and the second lowest income quintile, with a stabilising effect for the families situated in the mid-high part of the income distribution. Inequality in gross market income has generally been stable despite the significant shocks introduced by COVID-19. The relative positive outcome of the market income change is largely due to the flat rate wage subsidy for part-time and long-term casual workers, who could receive an income exceeding their usual wage.

On the disposable income side, we observe an increase in the living standards for most households except the ones in the richest income quintile since February although the increase is most evident in the bottom quintile similar to the change in the gross income. The increase in the bottom quintile can be mostly attributed to the policy response, which also negates the income shock effects for other quintiles. Due to the significant rise in the disposable income in the bottom quintile, Australia has experienced a rapid drop in income inequality, reducing the Gini in equalised disposable income from 0.33 to just above 0.30.

Our analysis also suggests that, without government interventions, both gross market and disposable incomes would have plunged considerably, with severe consequences in terms of raising income inequality and poverty levels. In absence of any policy response, the poverty rate would have jumped in the range of 3-12 percentage point compared with the pre-COVID period.

The temporary measures introduced (wage subsidy, increased benefit, free childcare) have successfully countered the shock, and in some metrics such as income inequality and poverty, dominated the impact of COVID-19, at least in short-term. Some individual policy measures, such as childcare and wage subsidy, are also found to have a similar impact compared with the findings from other countries with similar measures. While our analyses do not imply the measures have sheltered the shock for every household in need, the overall change in the income distribution does suggest the measures are generally progressive, protecting those who are on average more vulnerable.

Our results also suggest the change or the lack of change in the income distribution since the initial COVID-19 outbreak is heavily dependent on the temporary measures. This raises the question of what might happen should these policies be withdrawn. The previous global financial crisis has shown that countries which exceedingly rely on stimulus measures may experience issues with increased public debt, which may affect the sustainability of welfare schemes in the long term. Additionally, the presence of temporary measures may induce shifts in the labour market equilibrium in the long-term. These are relevant questions for researchers to further investigate in the future.

**REFERENCES**

ACOSS. (2020). *Poverty in Australia 2020 - part 1: overview.* http://povertyandinequality.acoss.org.au/wp-content/uploads/2020/02/Poverty-in-Australia-2020_Part-1_Overview.pdf

Australian Bureau of Statistics. (2019). *6553.0 - Survey of Income and Housing, User Guide, Australia, 2017-18.* https://www.abs.gov.au/ausstats/abs@.nsf/Lookup/6553.0main+features12017-18

Australian Bureau of Statistics. (2020a). *6160.0.55.001 - Weekly Payroll Jobs and Wages in Australia.* https://www.abs.gov.au/AUSSTATS/abs@.nsf/allprimarymainfeatures/C4682792DAAB8C55CA2585510005C748?opendocument

Australian Bureau of Statistics. (2020b). *6602.0 - Microdata: Longitudinal Labour Force, Australia.* https://www.abs.gov.au/AUSSTATS/abs@.nsf/Latestproducts/6602.0Main Features20Australia?opendocument&tabname=Summary&prodno=6602.0&issue=Australia&num=&view=

Australian Super. (2020). *Performance vs Benchmarks Quarterly.* https://www.australiansuper.com/api/graphs/quarterlyrates/table/download/super

Australian Treasury. (2020). *The JobKeeper Payment: Three-month review.* https://treasury.gov.au/sites/default/files/2020-07/jobkeeper-review-2020_0.pdf

Baker, S., Bloom, N., Davis, S., & Terry, S. (2020). COVID-Induced Economic Uncertainty. *National Bureau of Economic Research, Working Pa.* https://doi.org/10.3386/w26983

Baldacci, E., De Mello, L., & Inchauste, G. (2002). *Financial crises, poverty, and income distribution* (Issues 2002–2004). International Monetary Fund.

Bańbura, M., Giannone, D., Modugno, M., & Reichlin, L. (2013). Now-casting and the real-time data flow. *Handbook of Economic Forecasting*, *2*, 195–237. https://doi.org/10.1016/B978-0-444-53683-9.00004-9

Bargain, O., & Callan, T. (2010). Analysing the effects of tax-benefit reforms on income distribution: A decomposition approach. *Journal of Economic Inequality*, *8*(1), 1–21. https://doi.org/10.1007/s10888-008-9101-4

Bargain, O., Callan, T., Doorley, K., & Keane, C. (2017). Changes in Income Distributions and the Role of Tax-benefit Policy During the Great Recession: An International Perspective. *Fiscal Studies*, 1–27. https://doi.org/10.1111/1475-5890.12113

Beirne, K., Doorley, K., Regan, M., Roantree, B., & Tuda, D. (2020). The potential costs and distributional effect of COVID-19 related unemployment in Ireland. *Budget Perspectives*, *2021*.

Bok, B., Caratelli, D., Giannone, D., Sbordone, A. M., & Tambalotti, A. (2018). Macroeconomic Nowcasting and Forecasting with Big Data. *Annual Review of Economics*, *10*(1), 615–643.

https://doi.org/10.1146/annurev-economics-080217-053214

Bonaccorsi, G., Pierri, F., Cinelli, M., Flori, A., Galeazzi, A., Porcelli, F., Schmidt, A. L., Valensise, C. M., Scala, A., Quattrociocchi, W., & Pammolli, F. (2020). Economic and social consequences of human mobility restrictions under COVID-19. *Proceedings of the National Academy of Sciences of the United States of America*, *117*(27), 15530–15535. https://doi.org/10.1073/pnas.2007658117

Brewer, M., & Gardiner, L. (2020). The initial impact of COVID-19 and policy responses on household incomes. *Oxford Review of Economic Policy*.

Bronka, P., Collado, D., & Richiardi, M. (2020). *The COVID-19 Crisis Response Helps the Poor: the Distributional and Budgetary Consequences of the UK lock-down*. EUROMOD at the Institute for Social and Economic Research.

Centrelink. (2020). *Payments and services during coronavirus (COVID-19)*. https://www.servicesaustralia.gov.au/individuals/subjects/payments-and-services-during-coronavirus-covid-19

Cheong, K. S. (2001). Economic Crisis and Income Inequality in Korea. *Asian Economic Journal*, *15*(1), 39–60.

Chetty, R., Friedman, J. N., Hendren, N., & Stepner, M. (2020). How Did COVID-19 and Stabilisation Policies Affect Spending and Employment? A New Real-Time Economic Tracker Based on Private Sector Data. *National Bureau of Economic Research, Working Paper*, *27431*. https://doi.org/10.1017/CBO9781107415324.004

De Beer, P. D. E. (2012). Earnings and income inequality in the EU during the crisis. *International Labour Review*, *151*(4).

DiNardo, J., Fortin, N. M., & Lemieux, T. (1996). Labor Market Institutions and the Distribution of Wages, 1973-1992: A Semiparametric Approach. *Econometrica*, *64*(5), 1001–1044. https://doi.org/10.2307/2171954

Figari, F., & Fiorio, C. V. (2020). Welfare resilience in the immediate aftermath of the covid-19 outbreak in italy. *EUROMOD at the Institute for Social and Economic Research, Tech. Rep.*

Giannone, D., Reichlin, L., & Small, D. (2008). Nowcasting: The real-time informational content of macroeconomic data. *Journal of Monetary Economics*, *55*(4), 665–676. https://doi.org/10.1016/j.jmoneco.2008.05.010

Li, J., La, H. A., & Sologon, D. M. (2020). Policy, Demography, and Market Income Volatility: What Shaped Income Distribution and Inequality in Australia Between 2002 and 2016? *Review of Income and Wealth*, *0*, 1–26. https://doi.org/10.1111/roiw.12467

Li, J., & O'Donoghue., C. (2014). Evaluating binary alignment methods in microsimulation models. *Journal of Artificial Societies and Social Simulation*, *17*(1), 15.

Matsaganis, M., & Leventi, C. (2014). The Distributional Impact of Austerity and the Recession in Southern Europe. In *South European Society and Politics* (Vol. 19, Issue 3, pp. 393–412). Taylor & Francis. https://doi.org/10.1080/13608746.2014.947700

Mckibbin, W., & Fernando, R. (2020). *The Global Macroeconomic Impacts of COVID-19: Seven Scenarios* (Vol. 2, Issue 4).

National Skill Commission. (2020). *A snapshot in time - The Australian labour market and COVID-19*. https://www.nationalskillscommission.gov.au/sites/default/files/2020-06/NSC_a_snapshot_in_time_report.pdf

Navicke, J., Rastrigina, O., & Sutherland, H. (2014). Nowcasting Indicators of Poverty Risk in the European Union: A Microsimulation Approach. *Social Indicators Research*, *119*(1), 101–119. https://doi.org/10.1007/s11205-013-0491-8

O'Donoghue, C., Sologon, D. M., Kyzyma, I., & McHale, J. (2020). Modelling the distributional impact of the Covid-19 crisis. *Fiscal Studies*, *41*(2), 321–336.

**APPENDIX**

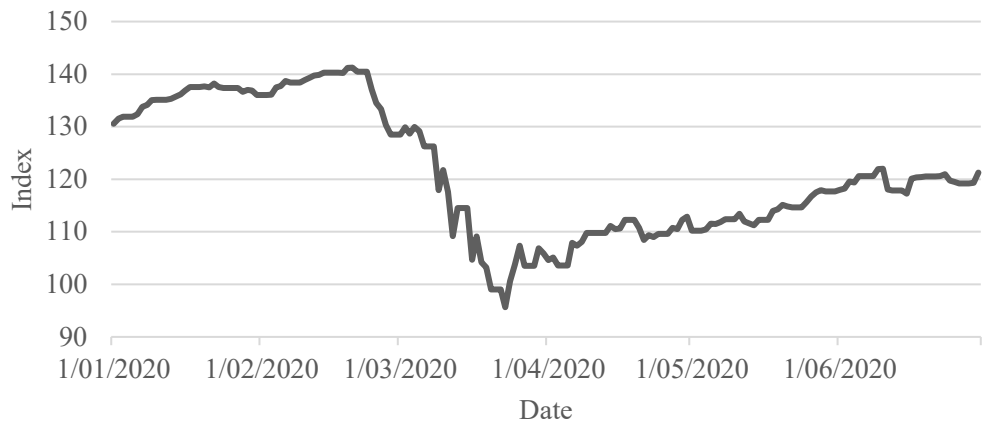

Figure A: Investment Performance (Index) of the default portfolio allocation (Balanced) of the Australian Super

Table B: Variables used in estimating the propensity of being employed at least once since February 2020

| | |
|---|---|
| Dependent variable | Whether one has been employed at least once since February 2020 conditional on currently not in the labour force |
| Independent variables used | Age, age squared, education, locally born, number of children between age 0 and 4, number of children aged between 5 and 14, state of residence |
| Sample | Observations who are currently out of the labour force. Estimated separately for male and female, and each wave of LLFS |

Table C: Variables used in estimating the propensity of remaining in employment conditional on previously employed

| | |
|---|---|
| Dependent variable | Whether one remains in employment |
| Independent variables used | Age, age squared, education, locally born, number of children between age 0 and 4, number of children aged between 5 and 14, state of residence current employment type, industry, occupation, number of usual hours worked, and number of jobs from the previous month |
| Sample | Observations who were currently employed in the previous month. Estimated separately for male and female, and each wave of LLFS |

Table D Changes in Number of Employed by Industry (Original)

|  | Feb 2020 (unit: 1000) | May 2020 (unit: 1000) | Change (%) |
|---|---|---|---|
| Agriculture, Forestry and Fishing | 337.1 | 361.3 | 7.2% |
| Mining | 238.5 | 227.3 | -4.7% |
| Manufacturing | 908.9 | 864.7 | -4.9% |
| Electricity, Gas, Water and Waste Services | 136.1 | 168.5 | 23.8% |
| Construction | 1182.1 | 1179.6 | -0.2% |
| Wholesale Trade | 385.7 | 389.3 | 1.0% |
| Retail Trade | 1261.4 | 1182.8 | -6.2% |
| Accommodation and Food Services | 930.5 | 654.3 | -29.7% |
| Transport, Postal and Warehousing | 666.4 | 571.8 | -14.2% |
| Information Media and Telecommunications | 211.6 | 187.9 | -11.2% |
| Financial and Insurance Services | 474.5 | 489.8 | 3.2% |
| Rental, Hiring and Real Estate Services | 213.7 | 219.9 | 2.9% |
| Professional, Scientific and Technical Services | 1171.9 | 1108.9 | -5.4% |
| Administrative and Support Services | 449.9 | 392.6 | -12.7% |
| Public Administration and Safety | 828.5 | 846.3 | 2.1% |
| Education and Training | 1096.6 | 1035.8 | -5.5% |
| Health Care and Social Assistance | 1798.3 | 1731.2 | -3.7% |
| Arts and Recreation Services | 251.9 | 160.1 | -36.5% |
| Other Services | 492.9 | 442.6 | -10.2% |
| Total Employed | 13036.7 | 12214.7 | -6.3% |

Source ABS Labour Force 6291.0.55.003 - Labour Force, Australia, Detailed, Quarterly, May 2020